\documentclass[
 aps,
 amsmath,amssymb,
reprint,
author-year,
superscriptaddress,longbibliography
]{revtex4-2}

\usepackage{graphicx}
\usepackage{bm}
\usepackage{upgreek}

\setcitestyle{super,open={},close={}}

\usepackage{hyperref}
\hypersetup{colorlinks = true,citecolor = blue,urlcolor = black}

\begin{document}

\graphicspath{{./Figures/}}

\title{{Emergent probability fluxes in confined microbial navigation}}

\author{Jan Cammann}
\thanks{equal contribution}
\affiliation{Interdisciplinary Centre for Mathematical Modelling and Department of Mathematical Sciences, Loughborough University, Loughborough, Leicestershire LE11 3TU, United Kingdom}
\affiliation{Max Planck Institute for Dynamics and Self-Organization, Am Fa{\ss}berg 17, 37077 G\"ottingen, Germany}
\author{Fabian Jan Schwarzendahl}
\thanks{equal contribution}
\affiliation{                    
  Institut f\"ur Theoretische Physik II: Weiche Materie, Heinrich-Heine-Universit\"at D\"usseldorf, 40225 D\"usseldorf, Germany}
\affiliation{Department of Physics, University of California, Merced, 5200 N. Lake Road, Merced, California 95343, USA}
\affiliation{Max Planck Institute for Dynamics and Self-Organization, Am Fa{\ss}berg 17, 37077 G\"ottingen, Germany}
\author{Tanya Ostapenko}
\affiliation{Max Planck Institute for Dynamics and Self-Organization, Am Fa{\ss}berg 17, 37077 G\"ottingen, Germany}
\author{Danylo Lavrentovich}
\affiliation{Max Planck Institute for Dynamics and Self-Organization, Am Fa{\ss}berg 17, 37077 G\"ottingen, Germany}
\author{Oliver B\"{a}umchen}
\affiliation{Max Planck Institute for Dynamics and Self-Organization, Am Fa{\ss}berg 17, 37077 G\"ottingen, Germany}
\affiliation{Experimental Physics V, University of Bayreuth, Universit\"atsstr.\ 30, 95447 Bayreuth, Germany}
\author{Marco G. Mazza}
\thanks{correspondence to m.g.mazza@lboro.ac.uk}

\affiliation{Interdisciplinary Centre for Mathematical Modelling and Department of Mathematical Sciences, Loughborough University, Loughborough, Leicestershire LE11 3TU, United Kingdom}
\affiliation{Max Planck Institute for Dynamics and Self-Organization, Am Fa{\ss}berg 17, 37077 G\"ottingen, Germany}

\date{\today}

\maketitle

\noindent
\textbf{When the motion of a motile cell is observed closely, it appears erratic, and yet the combination of nonequilibrium forces and surfaces can produce striking examples of organization in microbial systems.
While {most of our} current understanding is based on bulk systems or idealized geometries, it remains elusive how and at which length scale self-organization emerges in complex geometries. Here, using experiments, analytical and numerical calculations we study the motion of motile cells under controlled microfluidic conditions, and demonstrate that probability flux loops organize active motion even at the level of a single cell {exploring an isolated compartment of nontrivial geometry}. 
By accounting for the interplay of activity and interfacial forces, we find that the boundary's curvature determines the nonequilibrium probability fluxes of the motion.
We theoretically predict a universal relation between fluxes and global geometric properties that is directly confirmed by experiments.
Our findings open the possibility to decipher the most probable trajectories of motile cells and {may enable the design of geometries guiding their time-averaged motion}.}

\begin{center}
Significance Statement\\ 
\end{center}
\textbf{Motile microorganisms commonly live in porous media comprising microhabitats filled with interfaces of complex shape. On such small scales the interactions with these interfaces, rather than external gradients, dominate their motion in the search for favorable living conditions.
We demonstrate with experiments and theory that the geometry of confining interfaces shapes the topology of the most likely, average trajectory,
leading to directed fluxes of probability that are not exclusively localized at the near-wall region. 
Employing this principle allows to actively shape a microbe's average direction of movement, which could be of use in the design of topological transport mechanisms for microfluidic environments.
}

The  presence of hidden order and regularities in living systems ---seemingly intractable from the point of view of mathematics and physics--- was first intuited by Erwin Schr\"odinger~\cite{schroedinger1944}. 
More recently, there has been a renaissance of discoveries of physical principles governing living matter~\cite{vicsekPR2012,marchettiRMP2013}. In systems of motile microorganisms activity and geometry often conspire to create organized collective states\cite{riedelSC2005,sokolovPRL2012,woodhousePRL2012,wiolandPRL2013,grossmannEPJST2015,lushiPNAS2014,wiolandNatPhys2016,wiolandNJP2016,beppuSM2017,theillardSM2017,frangipaneNatComm2019,Wensink2014}. The robustness of these results invites the question: at what level such order starts emerging?  Specifically, is there a lower bound in either the number of participating microbes or the available space for regularities to affect the activity of the cells?
While the dynamics of active microswimmers, which propel themselves, e.g., by the periodic beating of one or multiple flagella~\cite{wan2016,boeddeker2020}, are often studied in idealized bulk situations, these microorganisms live in the proximity of interfaces, prosper in wet soil, inhabit porous rocks, and generally encounter complex boundaries regularly~\cite{ghiorseAdvApplMB1988,HeijdenEcolLett2008,gaddMicroBiol2010}. When colliding with a solid boundary, a microswimmer may interact with it through hydrodynamic interactions mediated by the fluid~\cite{polinSci2009,continoPRL2015}, through steric interactions~\cite{kantsler2013} or a combination of the two~\cite{elgetiEPJST2016, ElizabethHulme2008}. Studies of puller-type microswimmers like \textit{Chlamydomonas reinhardtii} suggest that for such microswimmers steric interactions dominate the dynamics in the presence of such boundaries~\cite{kantsler2013,ledesmaPRL2013,capriniSM2018,ostapenkoPRL2018}.  \textit{C.~reinhardtii} naturally lives in wet soil~\cite{harris2009}, an environment dominated by fluid-solid interfaces. A common observation is that such microswimmers are more likely to be found at or in close proximity of the boundary rather than within the interior~\cite{wysockiPRE2015,ibrahimEPJST2016,Elgeti2013,Elgeti2015}, showing a preference for regions of high wall curvature \cite{wysockiPRE2015,ostapenkoPRL2018, Fily2014}.
\begin{figure*}
    \centering
    \includegraphics[width=\textwidth]{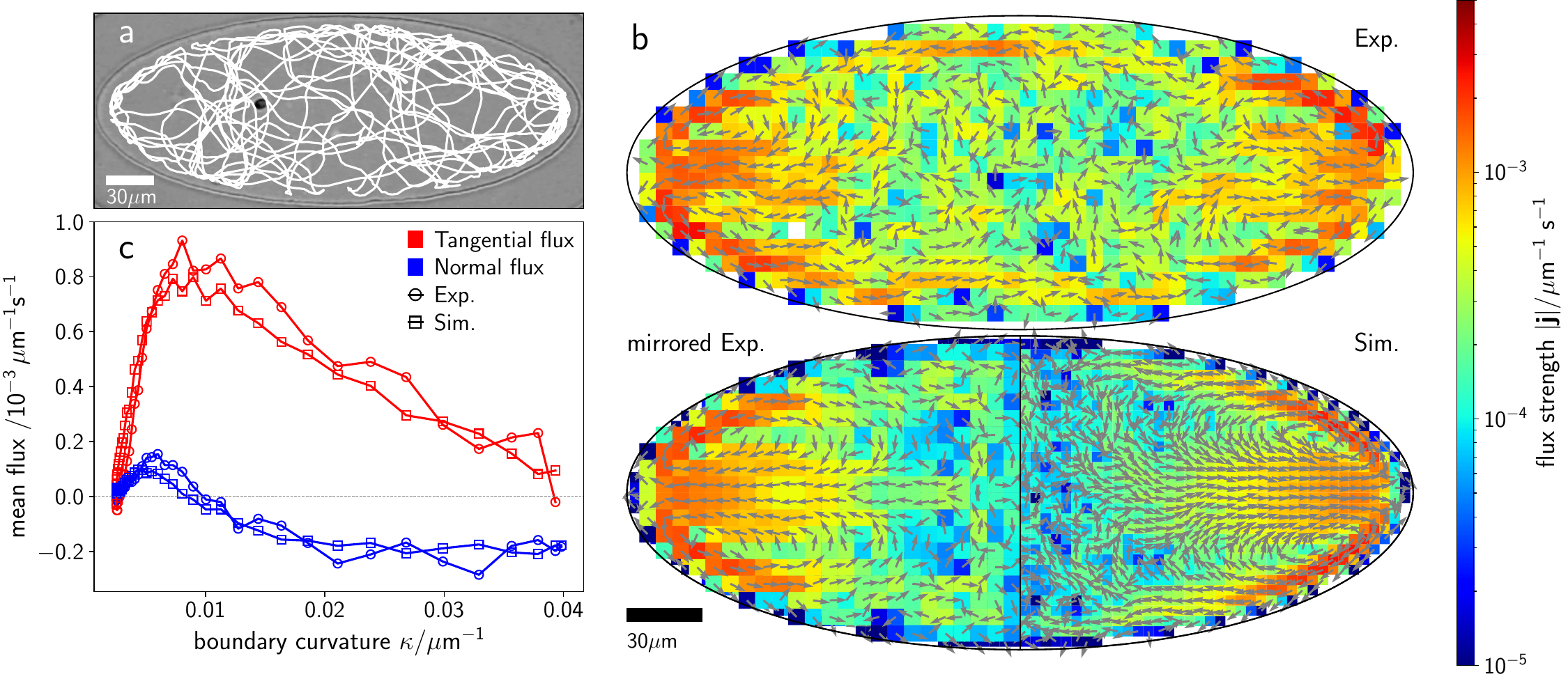}
    \caption{Cell trajectory and emerging nonequilibrium fluxes inside an elliptical compartment. 
    (a) Optical micrograph of a single \textit{C.~reinhardtii} cell contained in a quasi-two-dimensional (2D) elliptical compartment with a major semi-axis of $157\,\upmu$m and a minor semi-axis of $63\,\upmu$m. A representative section of the fully tracked cell trajectory is overlaid in white. The cell is visible at the left of the center. (b) Steady-state nonequilibrium fluxes with arrows indicating their direction and their strength encoded with color. The fluxes obtained by direct analysis of the experimental trajectories are shown on the top. Considering the system's symmetry we can increase the available statistics by mirroring along the major and minor axes of the ellipse and thus reducing the noise in the fluxes (bottom left). The fluxes resulting from Brownian dynamics simulations (see main text and SI for details on our model) are shown on the bottom right. (c) Steady-state fluxes in a strip within $20\,\upmu$m from the wall in tangential (red) and normal (blue) directions comparing experiments (circles) and simulations (rectangles). Positive tangential fluxes denote fluxes towards lower curvature; positive normal fluxes denote the direction away from the wall.}
    \label{fig:nonequilibrium_flux}
\end{figure*}

Here, we report experiments on a single \textit{C.~reinhardtii} cell confined in a compartment with varying boundary curvature, analytical calculations and simulations that model the microswimmer as an asymmetric dumbbell undergoing active Brownian motion and interacting sterically with the compartment wall. 
We find that when the compartment's boundary exhibits non-constant curvature, the accessible space is partitioned by loops of probability flux that direct and organize the cell's motion. 
This finding becomes evident when the nonequilibrium fluxes are extracted from the analysis of experimental and simulation data, and also estimated through the Fokker--Planck description of our system. The cell's front-back asymmetry results in a torque, that in turn causes an active reorientation when interacting with the wall. Such shape asymmetry proves crucial for the quantitative agreement of experiments and simulations making steric effects dominant over hydrodynamics in situations of strong confinement, for such flagellated microorganisms.
We show that the loops of probability flux are directly linked to the boundary's gradient of curvature, and can be quantified by a dimensionless number $\chi$ accounting for the relevant ratio of length scales.

We conduct experiments on single biflagellated \textit{C.~reinhardtii} cells, wild-type strain SAG 11-32b, confined within a quasi-two-dimensional microfluidic compartment in the absence of any in- and outlets. We employ optical bright-field microscopy and particle tracking techniques to extract cell trajectories over extended time periods~\cite{ostapenkoPRL2018}. Experiments were performed in elliptical compartments with eccentricities varying in the range $0.5-0.9$ and accessible areas of $(7 - 200)\times10^{3}$\,$\upmu$m$^2$. The height of all compartments was approximately 22\,$\upmu$m, i.e.\ slightly larger than one cell diameter, comprising cell body and flagella, such that the cell's motion was confined in two dimensions (see \emph{Methods}). 

We statistically investigate the average motion by tracking the cell's position over long trajectories. Figure \ref{fig:nonequilibrium_flux}a shows the overlaid positions of a single \textit{C.~reinhardtii} cell confined within a sealed elliptical compartment. While a certain preference to travel alongside the boundary is noticeable, frequent excursions across the microfluidic microhabitat are evident. 
Periods of time when the cell travels along the boundary are interrupted by rapid reorientations, crossing of the compartments, or curved arcs.  
The short-time motion of the cell is influenced by the stochasticity associated to the biological motors powering the flagella and their coordination~\cite{polinSci2009,geyer2013,wan2016}, and the small-scale hydrodynamic fluctuations of the fluid stirred by the swimming cell.

To investigate the nonequilibrium dynamics of the cell's motion, we compute the trajectory transition rates as follows. 
Starting from the cell's trajectories obtained via particle tracking, we divide the space in the compartment with a square grid and determine the probabilities to traverse a certain box in either direction by averaging over all passages of the trajectory on that box. By counting the directed transitions between adjacent boxes $\alpha$ and $\beta$ we can extract the net transition rates $w_{\alpha,\beta}$ from which the $x$ and $y$ components of the flux $\bm{j}$ are computed (see SI Sec.~I) \cite{ziaJSM2007,battleScience2016}.
This analysis reveals loops in the experimental probability fluxes, underlying the run-and-tumble-like  motion \cite{polinSci2009} of the cell within the compartment (Fig.~\ref{fig:nonequilibrium_flux}b).
Because of the symmetry of the system and the absence of any preferential direction in the cell's motion, the loops are symmetrically placed within the ellipse. The fluxes form a strong equatorial current pointing directly toward the ellipse's apices (the two points on the boundary with the highest curvature) and are then redirected along the boundary away from the apices. These loops are a manifestation of the inherent nonequilibrium nature of the cell's active motion, that explicitly breaks the invariance of the dynamics between a transition from one state to another and the inverse transition. 

Numerical simulations of our system lend insight into the properties of these flux loops.
We employ a minimal mechanistic model of \textit{C.~reinhardtii} cells as asymmetric dumbbells (see Fig.~S1),  {representing the characteristic fore-aft asymmetry of the cell's  body and of a larger space spanned by the stroke-averaged flagella}~\cite{robertsJFM2002,robertsBiolBullet2006,wysockiPRE2015}.
{Note that other microswimmer geometries produce fluxes as well, though without matching experiments quantitatively (see SI Sec.~V).}
The translational dynamics of the dumbbell are governed by an overdamped Langevin equation: $\frac{ \text{d} \bm{r} }{ \text{d} t } =  v_0 \bm{e} + \mu_w \bm{F}_w + \bm{\eta}$,
where $\bm{r}$ is the position of the geometric center of the dumbbell, $v_0$ is the self-propulsion velocity, $\bm{e}$ is the orientation of the dumbbell, $\mu_w$ is the mobility and $\bm{F}_w$ is the force stemming from steric wall interactions. Furthermore, $\bm{\eta}$ is a Gaussian white noise to account for translational diffusion with  coefficient $D_T = k_B T \mu_w$. 
The orientational dynamics are governed by the equation: $\frac{ \text{d} \bm{e} }{ \text{d} t } =  (  \bm{T}_w / \tau_w + \bm{\xi}) \times  \bm{e}$,
where $\bm{T}_w$ is the torque acting at the wall, $\tau_w$ is the rotational drag coefficient and $\bm{\xi}$ is a Gaussian white noise accounting for rotational diffusion with coefficient $D_R$. All parameters entering the equations of motion were either directly measured in our experiments or extracted from the literature (See \emph{Methods}).
\begin{figure}
    \centering
    \includegraphics[width=\columnwidth]{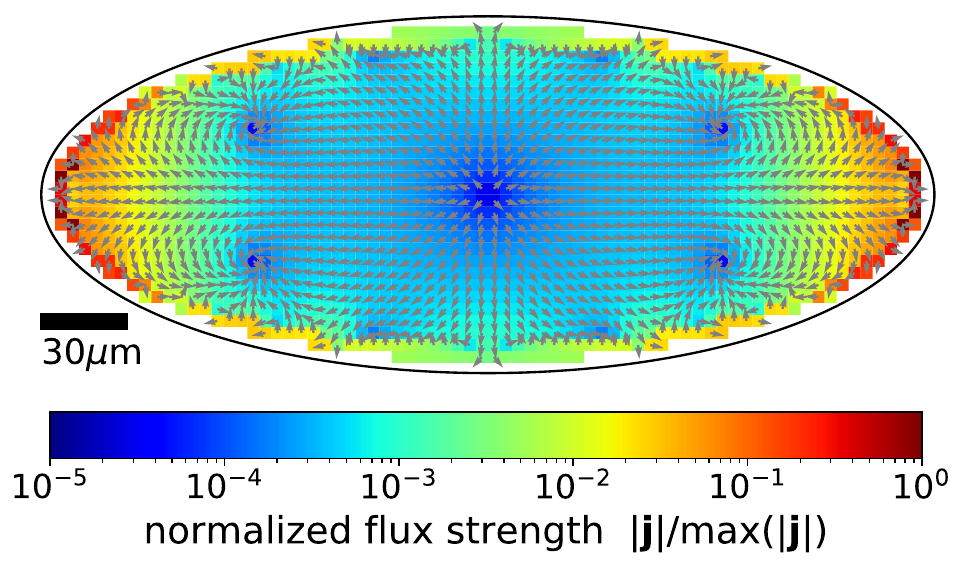}
    \caption{Topology of the nonequilibrium fluxes in an elliptical geometry. The structure of the flux loops is derived analytically from Eq.~\eqref{eq:noneqflux}. Nonuniform wall curvature induces four loops in the nonequilibrium fluxes, and strong equatorial fluxes pointing toward the ellipse's apices. Fluxes result from the interplay of the gradient of curvature and the polarization of active motion. See SI Sec.~II for details of the analytics.}
    \label{fig:theory_fluxes}
\end{figure}

Figure~\ref{fig:nonequilibrium_flux}b (bottom) also shows the results of our simulations for the probability flux for a single cell moving according to the above Langevin equations 
within an elliptical compartment with the same area and eccentricity as in the experiments. We find the analogous structure of flux loops emerging from the numerical results. Strong directional fluxes point from the central region of the compartment towards the apices and then move away along the boundary closing the flux loops. For any point on the compartment boundary the fluxes $\bm{j}$ can be decomposed in a component tangential and one normal to the boundary. We choose the positive normal direction to point into the compartment and the positive tangential direction to always point down the gradient of curvature. {The flux analysis reveals that in the bulk the cell is most likely to be on a path leading towards a region of high curvature} $\kappa$, as indicated by the negative normal fluxes in Fig.~\ref{fig:nonequilibrium_flux}c; {after colliding with the wall a smaller reorientation is needed to escape again at regions of low curvature, whereas escaping at high curvature requires larger turns, which are statistically less likely. This leads to a net statistical flux  along the wall towards regions of lower curvature} resulting in a rise of tangential flux with decreasing $\kappa$. 
The strong, outward equatorial fluxes (Fig.~\ref{fig:nonequilibrium_flux}b) meet the boundary and turn up or down along regions of lower curvature. 
At those turning points (close to the ellipse apices) the normal fluxes change sign, while the tangential flux become strongest. The quantitative agreement between experiments and simulations in Fig.~\ref{fig:nonequilibrium_flux}c {confirms the fundamental role of the force and torque applied by the boundary on the cell}.
\begin{figure*}
    \centering
    \includegraphics[width=\textwidth]{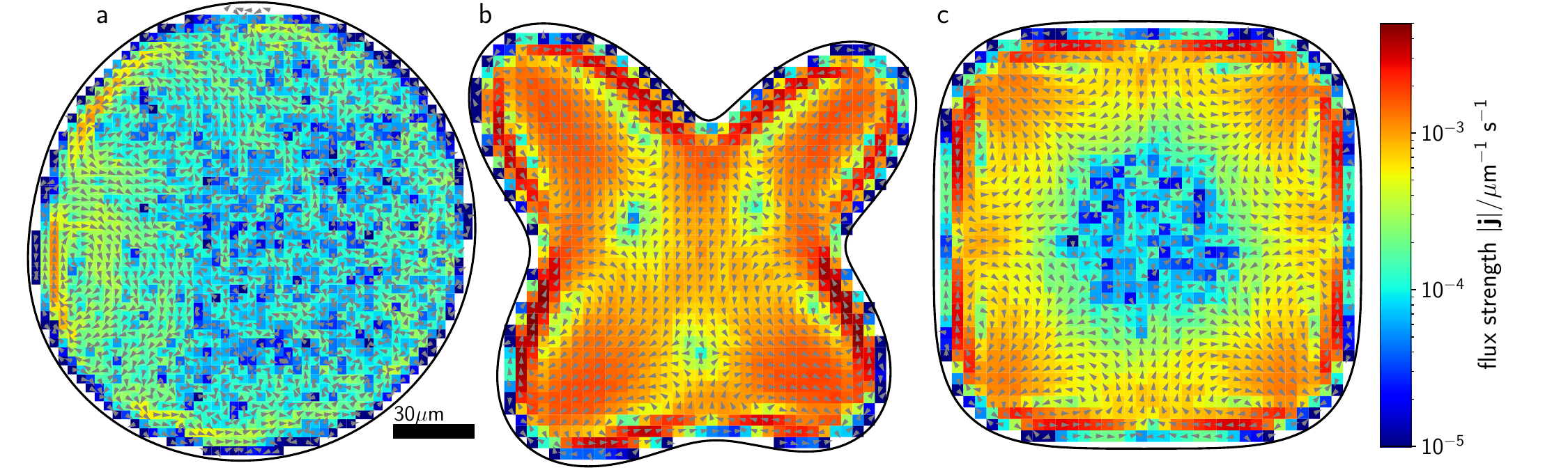}
    \caption{Complexity and topology of flux loops. Simulations of a model microswimmer in complex geometric confinement confirm the general topological features of the flux loops. The nonequilibrium fluxes are extracted from our active Brownian dynamics simulations. Fluxes are dominated by gradients of wall curvature. More compartment geometries are shown in Fig.~S4. The mathematical equations of the shown compartments are provided in SI Sec.~VI}
    \label{fig:compound}
\end{figure*}

For a more general understanding of the origin of the flux loops, we now turn to a continuum-mechanics approach of the probability flow by computing the Fokker--Planck equation for the system, which reads
\begin{align}
    \frac{\partial p}{\partial t} = &- \bm{\nabla} \cdot \Big(v_0 \bm{e} + \mu_w \bm{F}_w - D_T \bm{\nabla} \Big)p \nonumber \\
    & - \bm{e} \times \frac{\partial}{\partial \bm{e}} \cdot
    \left(\frac{1}{\tau_w} \bm{e} \times \bm{G}_w - D_R \bm{e} \times \frac{\partial}{\partial \bm{e}}\right)p\,,
    \label{eq:Fokker-planck-eq_main}
\end{align}
where the gorque $\bm{G}_w$ is a more convenient way to treat the torque $\bm{T}_w=\bm{e}\times\bm{G}_w$.
A moment expansion of the probability distribution function  $p=p(\bm{r},\bm{e},t)$ in terms of density $\rho(\bm{r})$ and polarization $\bm{P}(\bm{r})$ allows us to identify the nonequilibrium flux in position space $\bm{j}_{r}=v_0 \bm{P} + \mu_w \bm{F}_w \rho - D_T \bm{\nabla} \rho $. 
The probability flux $\bm{j}_{r}$ obeys a solenoidal condition $\bm{\nabla} \cdot \bm{j}_{r} = 0$ (see SI Sec.~II), which is  obeyed in both the experimental and simulated fluxes (Fig.~\ref{fig:nonequilibrium_flux}b). We can solve the dynamics of the probability loops by introducing a stream function $\bm{j}_{r}=\bm{\nabla}\times \bm{\psi}$, which satisfies the governing Poisson equation 
\begin{align}
    \Delta \psi = - \omega(\bm{F}_w,\bm{G}_w,\nabla\kappa)\,,
    \label{eq:noneqflux}
\end{align}
where the vorticity $\omega=\bm{\nabla}\times \bm{j}_{r}$ is a function of the forces $\bm{F}_w$ and gorques $\bm{G}_w$ exerted by the boundary on the swimming cell. Additionally, because of the anisotropic shape of the microswimmer, the vorticity crucially couples with the gradient of curvature $\nabla\kappa$.

The flux equation~\eqref{eq:noneqflux} with elliptical boundary conditions can be solved analytically (see SI Sec.~III). Figure~\ref{fig:theory_fluxes} shows the resulting fluxes.  
The qualitative features of the nonequilibrium fluxes found in experiments and simulations are reproduced by the solution of Eq.~(\ref{eq:noneqflux}). Four symmetrically-placed flux loops emerge in the region close to the apices. This fact points to a general connection between fluxes and boundary's curvature. We elucidate this relationship in the following.

We can deduct a quantitative relation of the fluxes from the above arguments. From our general arguments (see SI Sec.~II), the steady-state expression for the polarization in close proximity of the wall $\bm{P}_\mathrm{wall} \approx 
-{(2 \bm{\nabla} \cdot \bm{F}_w)^{-1}} \left( v_0 \bm{\nabla} + {\tau_w}^{-1} \bm{G}_w \right) \rho$. Recalling the definition of $\bm{j}_{r}$, and the fact that close to the boundary the probability density is proportional to the curvature $\rho=\alpha\kappa$~\cite{ostapenkoPRL2018}, i.e.\ the cell is more likely to spend more time in regions of high local curvature, we find that generally the flux depends on the curvature $\kappa$ and its gradient $\nabla\kappa$. In a circular domain, where the curvature is constant, symmetry prevents the emergence of fluxes (see SI Sec.~IV), even though the system is out of equilibrium. A local dependence of curvature breaks the symmetry and allows for nonzero fluxes. Thus, we expect the fluxes to depend on $\nabla\kappa$, but not directly on $\kappa$ as observed in our experiments and simulations.

We now generalize our arguments to more complex shapes by inferring the following effective rules: (\emph{i}) because flux loops are generated by curvature gradients, the number of flux loops equals the number of zero crossing of curvature gradient, $\nabla\kappa=0$; (\emph{ii}) the magnitude of a flux loop is proportional to the integrated change in curvature $\int \left|\frac{\partial \kappa}{\partial l}\right|\mathrm{d}l$ along a portion of the boundary with arc length $l$; (\emph{iii}) the number of stagnation points in the flux ($|\bm{j}_r|\approx0$) is at most one for every two flux loops. These predictions are in line with the topological structure of the fluxes for simulations of our model in compartments of growing complexity with multiple lobes and points of both positive and negative curvature, as shown in Fig.~\ref{fig:compound}.

\begin{figure*}
    \centering
    \includegraphics[width=0.7\textwidth]{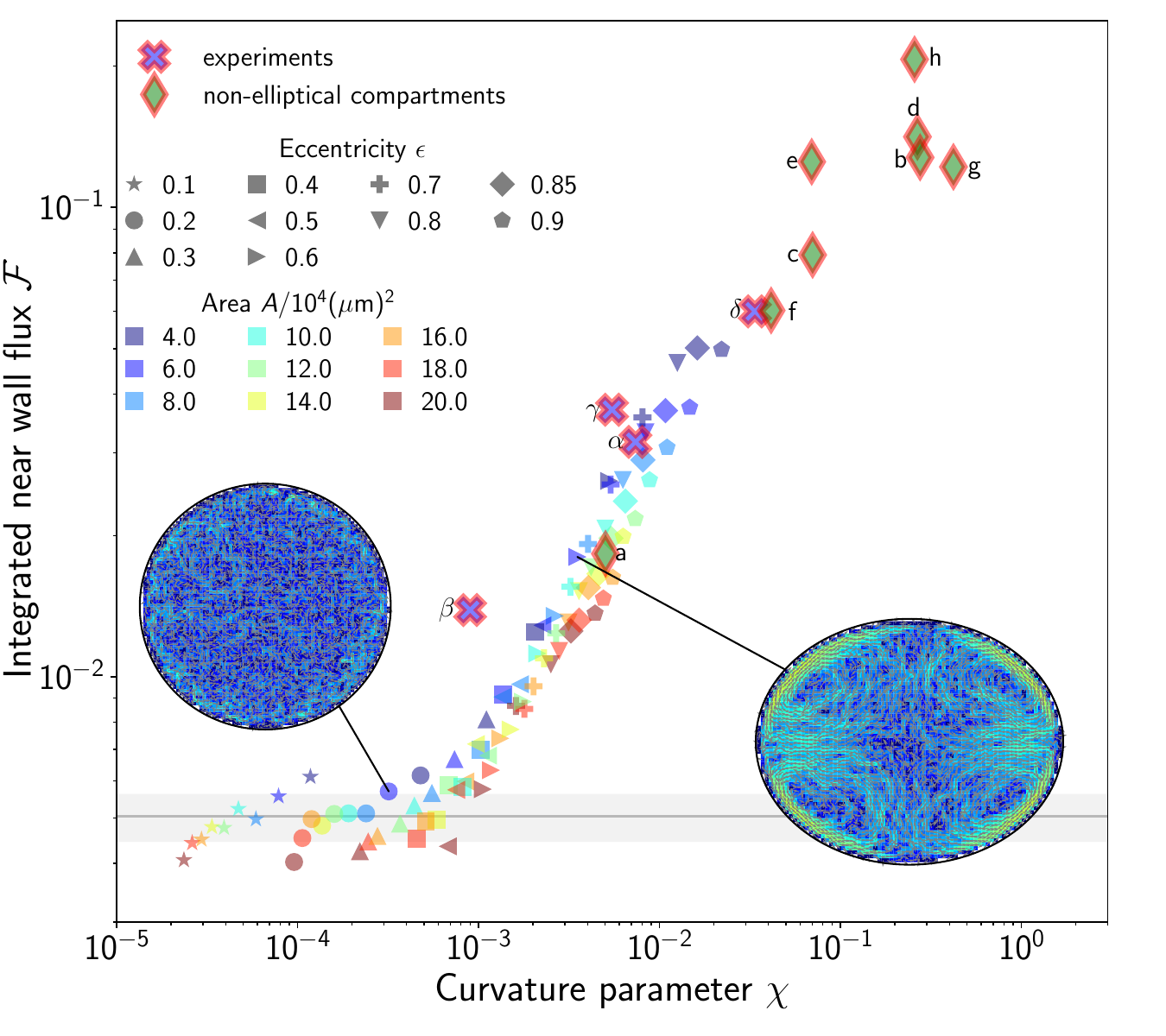}
    \caption{Relation between boundary curvature gradient and nonequilibrium fluxes. The integrated flux strength $\mathcal{F}$ depends solely on the curvature parameter $\chi$. Experiments and simulations for elliptical compartments with different eccentricities (solid symbols) and areas (color code) collapse on a master curve. The data gathered from experimentally recorded trajectories in elliptical compartments (framed crosses) as well as the simulations in complex shapes shown in Fig.~\ref{fig:compound} are also included here (framed diamonds). 
    Experiments were performed in compartments of eccentricities $\epsilon_\alpha\approx0.92$, $\epsilon_\beta\approx0.55$, $\epsilon_\gamma\approx0.55$, and $\epsilon_\delta\approx0.92$ with compartment areas $A_\alpha\approx14.23\times 10^4\,\mu$m$^2$, $A_\beta\approx19.09\times 10^4\,\mu$m$^2$, $A_\gamma\approx3.13\times 10^4\,\mu$m$^2$, and $A_\delta\approx3.11\times 10^4\,\mu$m$^2$.
    The crossover in the data at $\chi\approx 10^{-3}$ marks a transition from weak fluxes indistinguishable from noise with the available statistics to fully-developed flux loops.  
    Gray horizontal line and gray shaded area mark the average and standard deviation, respectively, for circular compartments with the same areas as the ellipses and the same amount of statistics. 
    }
    \label{fig:Hockey_stick}
\end{figure*}

Although Eq.~\eqref{eq:noneqflux} reveals the importance of forces acting at the boundaries, the nonequilibrium nature of the probability distribution imposes a nonlocal spatial distribution of the fluxes. 
It is then natural to consider the integral of the flux over an area of the compartment
\begin{align}
    \mathcal{F} \equiv \frac{1}{v_0}{\int_{S} |\bm{j}| \text{d} S},
    \label{eq:intfluxstr}
\end{align}
which gives an effective measure for the strength of the nonequilibrium fluxes within the region $S$  (we choose a strip of $20\,\upmu$m along the boundary where the fluxes are clearly distinguishable from statistical noise) and where self-propulsion velocity $v_0$ is used to nondimensionalize $\mathcal{F}$. 
Through $\bm{j}$,  $\mathcal{F}$ inherits the dependence on $\nabla\kappa\sim \tfrac{\partial \kappa}{\partial l}$. To capture the global characteristics of the geometry, we are naturally led to define the dimensionless number 
\begin{equation}
\chi\equiv A_c\frac{{\int{\left|\frac{\partial \kappa}{\partial l}\right|\mathrm{d}l}}}{\int \mathrm{d}l}
\end{equation}
comparing the  global change in curvature over the total perimeter $\int \mathrm{d}l$ of the boundary with the typical area of the cell $A_c$ calculated by squaring the swimmer's length 
(see \emph{Methods}).
We find that the integral flux strength $\mathcal{F}$ depends uniquely
on $\chi$ for both experiments and simulations of elliptical compartments with various areas $A$ and
eccentricities $\epsilon$ (Fig.~\ref{fig:Hockey_stick}). The crossover in the integrated fluxes at $\chi\approx 10^{-3}$ corresponds to the point at which the fluxes can be effectively distinguished from noise given the available statistics (in fact, to quantify the effect of statistical noise, we consider simulations in circular compartments --where fluxes are strictly absent-- and use the same amount of statistics as elliptical compartments; we find an average value $\mathcal{F}_\text{noise}=5\times10^{-3}$, marked  with a horizontal gray line; the shaded area marks the standard deviation derived from simulations of circular compartments with the same areas as the ellipses).

For values of scaled curvature gradients $\chi <10^{-3}$ the fluxes are rather weak, whereas for $\chi \geq 10^{-3}$ strong  fluxes emerge and exhibit closed loops such that their impact on the global dynamics within the compartment is much stronger. 
The near collapses of both experiments and simulations for all shapes confirms that the boundary's geometry determines the strength and shape of flux loops within the interior of the compartment.

Taken together, our results show that the boundary of the confining domain imposes a robust topology of loops of probability fluxes at the level of a single active cell. 
Our experimental and theoretical results demonstrate the intimate connection between the geometric properties of the boundary and the interior of a compartment confining a motile cell. The ensuing probability fluxes impose an organizing structure to the whole compartment's interior, that statistically guides  the cell's motion. Our study shows that \textit{C.~reinhardtii} cells are very efficient at exploring the available space, and that simple geometric features can leave an imprint on the cells' overall motion in their microhabitat. 

Harnessing the motion of microorganisms is a promising direction of technological development in active matter~\cite{dileonardoPNAS2010, ElizabethHulme2008}. Improved efficiency of micromachines will require a better understanding of how microbial cells navigate complex environments and interact with their boundaries. Inducing a statistical bias to directional motion, as a consequence of the nonequilibrium nature of the motion and confining boundaries, might in fact help producing efficient microdevices even at the scale of a single cell.

\noindent
\textbf{Acknowledgments}
The authors thank the G\"ottingen Algae Culture Collection (SAG) for providing the \textit{Chlamydomonas reinhardtii} wild-type strain SAG~11-32b. 
We gratefully acknowledge helpful discussions with Stephan Herminghaus. \\

\noindent
\textbf{Author Contributions}
M.G.M. designed research. J.C. and F.J.S. performed the theoretical work. J.C. carried out and analyzed the simulations. O.B. led the experiments. T.O. and D.L. performed them. J.C., F.J.S., T.O., O.B., and M.G.M. analyzed the data. All authors contributed to the discussions and the final version of the manuscript. 

\noindent
\textbf{Competing Interests}
The authors declare no competing interests.\\

\noindent
\textbf{Methods}

\noindent
\textbf{Cell Cultivation.} All experiments were performed using the wild-type \textit{Chlamydomonas reinhardtii} strain SAG 11-32b (provided by the G\"ottingen Algae Culture Collection, SAG).
The cells were cultivated axenically in Tris-Acetate-Phosphate (TAP) medium in a Memmert IPP110Plus incubator on a 12\,h day--12\,h night cycle. 
The temperature of the cell cultures in the incubator was kept at 24\,$^{\circ}$C during the day (white light, light intensity of about $3\times 10^{19}$ \,photons/m$^2$s in the center of the incubator) and reduced to a temperature of 22\,$^{\circ}$C during the night (no illumination), respectively. 
Prior to every experiment, about 50\,mL of cell culture was centrifuged (Eppendorf 5804R) for about 10 minutes at an acceleration of 100\,$g$ at ambient temperature.
About 40-45\,mL of the suspension was then removed and the remaining 5-10\,mL were allowed to relax for 90-120 minutes in the incubator to ensure that the cells have a sufficient amount of time to eventually regrow their flagella. 
Finally, this suspension was diluted with cell culture medium to very low cell concentrations in order to enhance the probability of capturing precisely one single cell per compartment.\\

\noindent
\textbf{Microfluidics.} The experimental chambers were composed of 2D arrays of stand-alone elliptical microfluidic compartments exhibiting a height of about 22\,$\upmu$m.
These compartment arrays were manufactured by employing soft lithography techniques using a curable elastomer (polydimethysiloxance, PDMS, Sylgard 184 Elastomer Kit, Dow Inc.) and master structures, which we produced by means of photolithography techniques in a cleanroom. 
Before filling the chambers with the cell culture, the PDMS-based microfluidic device and a glass microscope slide were treated with air plasma (Pico, Diener Electronic) for 30\, seconds to render their surfaces hydrophilic. 
A droplet of the diluted cell suspension was placed onto the feature side of the microfluidic device such that the compartment array was completely filled and, subsequently, the glass slide was placed atop and gently pressed to tightly seal the experimental compartment. {The confining surfaces at the top and bottom prevent the organism from turning out of the plane, enabling a 2D description.}\\

\noindent
\textbf{Microscopy.} For time-resolved cell imaging, we employed bright-field microscopy (Olympus IX-81 inverted microscope) in controlled light conditions (closed box).
During all experiments, the microfluidic compartment was illuminated using a narrow red-light bandpass filter (671\,nm, full-width-half-maximum FWHM=10\,nm) in order to exclude any photoactive response of the cell, including phototaxis\cite{berthold2008channelrhodopsin,foster1984rhodopsin} and light-induced adhesion to surfaces\cite{kreis2018adhesion,kreis2019}. 
A Canon 600D camera (at 25 frames per second, resolution: 1920$\times$1080\,pixels) was used to record videos of single cells swimming in isolated elliptical compartments for about 5--30\,minutes each. 
In order to increase the statistics substantially, these single-cell recordings were independently repeated 3--8 times for each chamber geometry.\\ 

\noindent
\textbf{Image Processing and Cell Tracking.} All videos were converted into sequences of 8-bit grayscale images with improved contrast using a custom-made \textsc{Matlab} algorithm.
The compartment boundaries were manually identified to restrict the cell tracking to the region available to the motion of the cell. 
Two-dimensional cell tracking was finally performed using \textsc{Matlab} based on the protocol developed by Crocker and Grier \cite{crocker1996methods}.\\

\noindent
\textbf{Numerical model and simulation parameters.}
The \textit{C.~reinhardtii} cells are modeled as asymmetric dumbbells {(see Fig.~S1) with a large sphere in front and a smaller sphere in the back, representing the fore-aft asymmetry of body and appendages}~\cite{robertsJFM2002,robertsBiolBullet2006}.
The equation of motion for the position $\bm{r}$ of the active dumbbell is given by 
\begin{align}\label{eq:pos_eom}
\frac{ \text{d} \bm{r} }{ \text{d} t } =  v_0 \bm{e} + \mu_w \bm{F}_w + \bm{\eta}.
\end{align}
Here, $v_0 = 100\,\upmu \text{m/s}$ is the self-propulsion speed of the cell, and
$\bm{\eta}$ is a Gaussian white noise with correlator $\langle \bm{\eta}(t) \bm{\eta}(t') \rangle = 2 k_\mathrm{B}T\mu_w \mathbf{1} \delta(t-t')$ and translational diffusion coefficient $k_\mathrm{B}T\mu_w = 250\,\upmu$m$^{2}/s$ (both numerical values are taken from \cite{ostapenkoPRL2018}). 
The term $\bm{F}_w$ accounts for steric wall interactions of the dumbbell and 
is computed by $\bm{F}_w = \bm{F}_1 + \bm{F}_2$ 
with $\bm{F}_{\alpha} = - \bm{\nabla} U_{\alpha}(r)$, $\alpha=1,2$, where $1$ and $2$ refers to the large and small sphere of the dumbbell, respectively. To compute the respective steric forces we use the Weeks--Chandler--Anderson potential~\cite{Weeks2003}
\begin{align}
    U_{\alpha} (d)/(k_\mathrm{B}T) =  4 \epsilon \left[ \left(\frac{a_{\alpha}}{d} \right)^{12}-  \left( \frac{a_{\alpha}}{d}\right)^6 \right] + \epsilon, 
    \label{eq:WCA}
\end{align}
if $d < 2^{1/6}a_{\alpha}$, and $0$ otherwise,                   
where $d$ is the distance of the sphere $\alpha\in \{1,2\}$ to the wall of the compartment.
The radii of the large (1) and small (2) circle of the dumbbell are $a_{1}=5\,\upmu$m, $a_2=2.5\,\upmu$m and we use $\epsilon=10$ to obtain a sufficiently strong screening (the values are taken from \cite{ostapenkoPRL2018}).

The orientation $\bm{e}$ is defined as the unit vector pointing from the small to the large circle of the dumbbell.
The equation of motion for the orientation is given by
\begin{align}\label{eq:orien_eom}
\frac{ \text{d} \bm{e} }{ \text{d} t } =  (  \bm{T}_w / \tau_w + \bm{\xi}) \times  \bm{e}.
\end{align}
Here $\bm{\xi}$ is a Gaussian white noise with correlator $\langle\bm{\xi}(t) \bm{\xi}(t') \rangle = \frac{2 k_\mathrm{B}T}{\tau_p} \mathbf{1} \delta(t-t')$ and rotational diffusion coefficient $\frac{\tau_p}{k_\mathrm{B}T} = 2\,\text{s}$ (see \cite{kantsler2013}). 
The torque acting at the wall is computed by $\bm{T}_w = \bm{T}_1 + \bm{T}_2$, where we use $\bm{T}_1 = (\bm{r}_1 - \bm{r}) \times \bm{F}_1 = l (\bm{e} \times \bm{F}_1)/2$,  $\bm{T}_2 = -l (\bm{e} \times \bm{F}_2)/2$, and $l=5\,\upmu$m.
For the shear time at the wall we use $\frac{\tau_w}{k_\mathrm{B}T} = 0.15\,\text{s}$ (see \cite{kantsler2013}).
In addition to Eq.~\eqref{eq:orien_eom} we included a run-and-tumble motion of the cell. Each tumbling event is instantaneous and the time between each tumbling event is sampled for an exponential distribution with mean $\frac{\tau_p}{k_\mathrm{B}T}$. The relative tumbling angle $\phi_{\mathrm{tumble}}$ is drawn from a Gaussian distribution with 
a standard deviation of 0.1 and a mean of $\tfrac{\pi}{2}$.


\clearpage

\onecolumngrid

\renewcommand{\thefigure}{S\arabic{figure}}
\renewcommand{\theequation}{S\arabic{equation}}
\setcounter{figure}{0}    
\setcounter{equation}{0}    

\begin{center}
    {\Large Supplementary Information}
\end{center}

\renewcommand{\thefigure}{S\arabic{figure}}
\renewcommand{\theequation}{S\arabic{equation}}

\section{Numerical computation of nonequilibrium fluxes}
\label{sec:flux:computation}
We numerically compute the nonequilibrium probability fluxes using a method introduced by Battle et al.~\cite{battleScience2016}. We divide positional space into equally sized square boxes $(i,j)$ of side $\Delta x$, where $i$ and $j$ denote the box's position in $x_1$ and $x_2$ direction.
From the recorded trajectories of the \emph{Chlamydomonas} cells we construct a time series $A(t_n)$ containing information about the cell's location $(i,j)_n$ (the box it resides in) at time $t_n$,
and the time $t_{n,n+1}$ spent in the state $(i,j)_n$ before the transition to the new state $(i,j)_{n+1}$ occurs. Quite generally, $A(t_n)$ in matrix form reads
\begin{align}
   A(t_n)= \left[\begin{array}{lll}
        (i,j)_1 & (i,j)_2 & ~t_{1,2}  \\
        (i,j)_2 & (i,j)_3 & ~t_{2,3}  \\\\
        &\quad\vdots\\\\
        (i,j)_{N-1} & (i,j)_N & ~t_{N-1, N}
    \end{array}\right]\,,
\end{align}
where the index $n$ indicates the discrete time steps, and $N$ is the total length of the time series.
Limited time resolution of the continuous trajectory $\bm{x}(t)\rightarrow(i,j)_n$ might lead to entries in $A$ where two successive states $(i,j)_n$ and $(i,j)_{n+1}$ do not correspond to adjacent boxes. In such cases, we determine the intermediate boxes via linear interpolation and insert them into $A(t_n)$, such that contiguous rows in $A(t_n)$ correspond to neighboring boxes.

The stochasticity of the system necessitates a large amount of statistics to identify significant fluxes. To maximize the amount of available data, the trajectories recorded experimentally were mirrored along the symmetry axes of the elliptical compartments, effectively quadrupling the amount of available trajectory data.
The transition rates $w_{(i,j),(k,l)}$ between boxes $(i,j)$ and $(k,l)$ can be calculated by counting all rows of $A(t_n)$ containing a transition from $(i,j)$ to $(k,l)$ and those that contain transitions in the opposite direction
\begin{align}
    w_{(i,j),(k,l)} = \frac{1}{t_\text{total}}\Big(N_{(i,j),(k,l)}-N_{(k,l),(i,j)} \Big)\,,
\end{align}
where $N_{(i,j),(k,l)}$ denotes the number of transitions from box $(i,j)$ to $(k,l)$, $N_{(k,l),(i,j)}$ the number of transitions in the opposite direction, and $t_\text{total}$ the total duration of the trajectory. 
The coarse-grained probability flux corresponding to box $(i,j)$ is then calculated as 
\begin{align}
    \bm{j}_{(i,j)} = \frac{1}{2\Delta x}\begin{pmatrix}
         w_{(i-1,j)(i,j)}+w_{(i,j)(i+1,j)}  \\
         w_{(i,j-1)(i,j)}+w_{(i,j)(i,j+1)} 
    \end{pmatrix}\,.
\end{align}
By bootstrapping the rows of $A(t_n)$ we can calculate statistical uncertainties of the coarse-grained flux $\bm{j}$, and by probing for correlations between consecutive rows one gains information about whether the system is Markovian or not.  For more details on such procedures the reader may be referred to the supplemental material of \cite{battleScience2016}.

{To ensure sufficient resolution of the experimental fluxes the long axis of the ellipse is divided into 50 bins. For the simulations, 80 bins along the long ellipse axis were used. Non-elliptical compartments were resolved at a similar scale. Using circular compartments (where fluxes are absent), we estimate the background noise due to the finiteness of statistics to be $ \mathcal{F}_\textrm{noise}\approx (5\pm0.6)\times 10^{-3}$ shown by a gray horizontal line in Fig.~4.}

\section{Analytical treatment}
\label{sec:analytical}

\begin{figure}[t]
    \centering
    \includegraphics[width=0.3\textwidth]{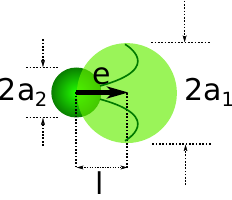}
    \caption{In simulations the swimmer is modelled as an asymmetric dumbbell of two spheres a distance $l$ apart. The back sphere represents the cell body and the front sphere models the stroke averaged shape of the flagella, which are beating fast enough, that on the relevant time scales for steric interactions they can be approximated as a solid sphere.}
    \label{fig:dumbbell_illustration}
\end{figure}

From the microscopic Eqs.~(5-7) we compute the Fokker--Planck equation for the probability $p=p(\bm{r},\bm{e},t)$ to find a \textit{C. reinhardtii} cell, which reads
\begin{align}
    \frac{\partial p}{\partial t} = &- \bm{\nabla} \cdot \Big(v_0 \bm{e} + \mu_w \bm{F_w} - D_T \bm{\nabla} \Big)p
    - \bm{e} \times \frac{\partial}{\partial \bm{e}} \cdot
    \left(\frac{1}{\tau_w} \bm{e} \times \bm{G}_w - D_R \bm{e} \times \frac{\partial}{\partial \bm{e}}\right)p.
    \label{eq:Fokker-planck-eq}
\end{align}
Here we used the approximation that both force $\bm{F}_w= \bm{F}_w(\bm{r})$ and gorque $G_w$, $\bm{T}_w= \bm{e} \times \bm{G}_w(\bm{r})$, only depend on the position $\bm{r}$. 
Equation~\ref{eq:Fokker-planck-eq}  can be written symbolically as
\begin{align}
    \frac{\partial p}{\partial t} = - \bm{\mathcal{L}} \cdot \bm{\mathcal{J}}
    \label{eq:FPeq_div_flux}
\end{align}
with the following definitions
\begin{align}
    & \bm{\mathcal{L}}= \begin{pmatrix}\bm{\nabla}\\\bm{e} \times \frac{\partial}{\partial \bm{e}}\end{pmatrix}\,,
    \label{eq:generaldiv}
    \\
    &\bm{\mathcal{J}}= \begin{pmatrix}
    v_0 \bm{e} + \mu_w \bm{F}_w - D_T \bm{\nabla}
   \\
    \frac{1}{\tau_w} \bm{e} \times \bm{G}_w - D_R \bm{e} \times \frac{\partial}{\partial \bm{e}}\end{pmatrix}p\,,
        \label{eq:generalJ}
\end{align}
for the operator $\mathcal{L}$ and probability flux $\mathcal{J}$. 

To make progress with Eq.~\eqref{eq:Fokker-planck-eq} we use a multipole expansion and compute equations for the density $\rho(\bm{r})= \int p(\bm{r},\bm{e},t) \mathrm{d} \bm{e}$ and polarization $\bm{P}(\bm{r}) = \int \bm{e} p(\bm{r},\bm{e},t) \mathrm{d} \bm{e}$, which read
\begin{align}
    \frac{\partial \rho}{\partial t} = 
    &-  \bm{\nabla} \cdot \left( v_0 \bm{P} + \mu_w \bm{F}_w \rho - D_T \bm{\nabla} \rho \right) + \bm{P} \cdot \bm{G}_w,
    \label{eq:densityfull}
    \\
    \frac{\partial \bm{P}}{\partial t} = 
    &-\frac{1}{2} \left( v_0 \bm{\nabla} 
    - \frac{1}{\tau_w} \bm{G}_w \right) \rho 
      - \mu_w \bm{\nabla} \cdot (\bm{F}_w \bm{P}) + D_T \bm{\nabla}^2 \bm{P} - D_R \bm{P}.
     \label{eq:polarizationfull}
\end{align}

To find the nonequilibrium flux of the density $\rho$, 
we now compute the orientational average of the probability flux Eq.\eqref{eq:generalJ}, which can be expressed in terms of the density and polarization
\begin{align}
    \int \bm{\mathcal{J}} \mathrm{d}^2\bm{e} = 
    \begin{pmatrix}
    v_0 \bm{P} + \mu_w \bm{F}_w \rho - D_T \bm{\nabla} \rho 
   \\
    \frac{1}{\tau_w} \bm{P} \times \bm{G}_w\end{pmatrix}
    = 
    \begin{pmatrix}
    \bm{j}_{r} \\\bm{j}_{e} 
    \end{pmatrix},
    \label{eq:fluxdef}
\end{align}
which defines the translational flux $\bm{j}_{r}$ and rotational flux $\bm{j}_{e}$.
The translational flux can be identified in Eq.\eqref{eq:densityfull}, which then simplifies to
\begin{align}
        \frac{\partial \rho}{\partial t} = 
    &-  \bm{\nabla} \cdot \bm{j}_{r} + \bm{P} \cdot \bm{G}_w.
    \label{eq:densityflux}
\end{align}
If we now assume a nonequilibrium steady state we arrive at
\begin{align}
      \bm{\nabla} \cdot \bm{j}_{r} = \bm{P} \cdot \bm{G}_w.
    \label{eq:noneqfluxAP}
\end{align}
Here, it is worth pointing out that $\bm{G}_w$ is only nonzero at the boundary.
Since the \textit{C.~reinhardtii} cell swims mostly parallel to the wall,  $\bm{P}$ is parallel to the wall; further, $\bm{G}_w$ is by definition normal to the wall; thus it follows
\begin{align}
      \bm{\nabla} \cdot \bm{j}_{r} = 0.
    \label{eq:noneqfluxdivfreee}
\end{align}
This condition states that the nonequilibrium fluxes are divergence-free. 
To determine the nonequilibrium fluxes we now use the vector-potential definition of stream function $\bm{j}_{r}=\nabla\times \bm{\psi}$, where the vector potential $\bm{\psi}\equiv  \begin{pmatrix}
     0\\
     0\\
     \psi
     \end{pmatrix}$, that is, in coordinates
     \begin{align}
     \begin{pmatrix}
     j_{r,x}\\
     j_{r,y}\\
     0
     \end{pmatrix}
     = \nabla \times
     \begin{pmatrix}
     0\\
     0\\
     \psi
     \end{pmatrix}.
    \label{eq:streamfunctiondef}
\end{align}
     We can find a governing equation for $\bm{\psi}$ by considering the vorticity $\omega= \partial_x j_{r,y} - \partial_y j_{r,x}$ and using Eq.~\eqref{eq:streamfunctiondef}.

This leads to a divergence-free $\bm{j}_{r}$ that is determined by the following Poisson equation
\begin{align}
    \Delta \psi = - \omega\,.
    \label{eq:vorticityPoisson}
\end{align}

The vorticity can be determined using the definition of the nonequilibrium flux in Eq.~\eqref{eq:fluxdef}
\begin{align}
     \begin{pmatrix}
     0\\
     0\\
     \omega
     \end{pmatrix}=
    \nabla \times \begin{pmatrix}
     j_{r,x}\\
     j_{r,y}\\
     0
     \end{pmatrix}=\nabla \times\bm{j}_{r} 
     = \nabla \times  ( v_0 \bm{P} + \mu_w \bm{F_w} \rho - D_T \bm{\nabla} \rho ).
    \label{eq:vorticityStart}
\end{align}
The last term on the right-hand side of Eq.~\eqref{eq:vorticityStart} vanishes identically;  the second term can be rewritten as
\begin{align}
    \mu_w \nabla \times \bm{F_w} \rho= 
    \mu_w ( \rho \nabla \times \bm{F_w}
    + \nabla \rho \times \bm{F_w})\,,
\end{align}
where the first term vanishes since $\bm{F_w}$ is the gradient of a potential. 
After these simplifications, the vorticity reads
\begin{align}
     \begin{pmatrix}
     0\\
     0\\
     \omega
     \end{pmatrix}=  v_0  \nabla \times \bm{P} + \mu_w \nabla \rho \times \bm{F_w}.
    \label{eq:vorticity1}
\end{align}
We now derive an expression for the curl of the polarization, $\nabla\times \bm{P}$.
In the steady state, the polarization equation reads (see Eq.~\eqref{eq:polarizationfull})
\begin{align}
    0 = 
    &-\frac{1}{2} \left( v_0 \bm{\nabla} 
    - \frac{1}{\tau_w} \bm{G}_w \right) \rho 
    - \mu_w \bm{\nabla} \cdot (\bm{F}_w \bm{P}) + D_T \bm{\nabla}^2 \bm{P} - D_R \bm{P}\,.
     \label{eq:polarizationfullAppendix}
\end{align}
Taking the curl of Eq.\eqref{eq:polarizationfullAppendix}, and neglecting translational diffusion gives
\begin{align}
    0 = 
    &
    \frac{1}{2 \tau_w} \nabla \times (\bm{G}_w  \rho)
    - \mu_w \bm{\nabla} \times( \bm{P} \nabla \cdot\bm{F}_w + \bm{F}_w \nabla \cdot \bm{P}) - D_R \nabla \times \bm{P}.
     \label{eq:polarizationfullApprox2}
\end{align}
Since the \textit{C.~reinhardtii} cell swims mostly parallel to the wall, we do not 
expect a large divergence of the polarization close to the boundary and thus we 
can  neglect the term $\bm{F}_w \nabla \cdot \bm{P}$ in Eq.~\eqref{eq:polarizationfullApprox2}; additionally, the term $\nabla(\nabla\cdot \bm{F}_w)$ is approximately parallel to the wall, and hence to $\bm{P}$. The curl of the polarization reads
\begin{align}
    \nabla \times \bm{P} = 
     \frac{\nabla \times (\bm{G}_w  \rho)}{ 2 \tau_w (\mu_w \bm{\nabla} \cdot \bm{F}_w  + D_R   )}\,.
     \label{eq:polarizationApproxAppendix}
\end{align}
Plugging Eq.~\eqref{eq:polarizationApproxAppendix} into Eq.~\eqref{eq:vorticity1} gives
\begin{align}
     \begin{pmatrix}
     0\\
     0\\
     \omega
     \end{pmatrix}=  v_0 \,
   \frac{\nabla \times (\bm{G}_w  \rho)}{ 2 \tau_w (\mu_w \bm{\nabla} \cdot \bm{F}_w  + D_R   )}  + \mu_w \nabla \rho \times \bm{F_w}.
    \label{eq:vorticity2}
\end{align}
Here, it is worth noting that all terms that lead to a vorticity in Eq.~\eqref{eq:vorticity2}  act only at the boundary of the compartment. 
From \cite{ostapenkoPRL2018} we know that the density at the wall approximately scales with 
the curvature $\rho_\mathrm{wall} \approx \alpha \kappa$, where $\kappa$ is the local curvature at the wall and $\alpha$ is a constant.
Furthermore, the gorque $\bm{G}_w$ is a gradient of a potential such that  Eq.~\eqref{eq:vorticity2} can be simplified to
\begin{align}
\omega =
v_0 \alpha \,\frac{G_{w,y}  \partial_x \kappa - G_{w,x}  \partial_y \kappa}{ 2 \tau_w (\mu_w \bm{\nabla} \cdot \bm{F}_w  + D_R   )}  + \mu_w (F_{w,y} \partial_x \kappa - F_{w,x} \partial_y \kappa ).
\label{eq:vorticityFinal}
\end{align}
Using Eq.~\eqref{eq:vorticityFinal} for the vorticity in Eq.~\eqref{eq:vorticityPoisson} gives
\begin{align}
    \Delta \psi =  -v_0 \alpha\, \frac{G_{w,y}  \partial_x \kappa - G_{w,x}  \partial_y \kappa}{ 2 \tau_w (\mu_w \bm{\nabla} \cdot \bm{F}_w  + D_R   )}  - \mu_w (F_{w,y} \partial_x \kappa - F_{w,x} \partial_y \kappa ),
    \label{eq:PoissonSolve}
\end{align}
which can be solved exactly (see next Section).

\section{Solution of Poisson equation}
The Green's function of the two-dimensional Poisson equation in an ellipse is given by\cite{liemertAMC2014} 
\begin{align}
    G(z,z_0) = -\frac{1}{2 \pi} \ln \frac{2 |z-z_0| }{ A+B }
    +\frac{1}{2 \pi} \sum_{k=0}^{\infty} \ln 
    \left|  
    \frac{4 q^{4k+1} [z^2 + (z_0^*)^2 ]  - 4 q^{2k} (1 + q^{4k+2}) z z_0^* + (A+B)^2 (1-q^{4k+2})^2}
    {4 q^{4k+3} (z^2 + z_0^2) - 4q^{2k+1} (1+ q^{4k+4}) zz_0 + (A+B)^2 (1- q^{4k+4})^2}
    \right|,
    \label{eq:GreensPoisson}
\end{align}
where we use the complex variables $z= x+ iy$ for the position and $z_0= x_0+ iy_0$ for the position of the source at $(x_0,y_0)$, $A$ and $B$ are the semi-major and semi-minor axes of the ellipse, respectively, and $q= (A-B)/(A+B)$. 

Formally the solution of the Poisson equation \eqref{eq:vorticityPoisson} is then given by the convolution (denoted with $*$)
of Eq.~\eqref{eq:vorticityFinal} and Eq.~\eqref{eq:GreensPoisson}, which reads
\begin{align}
    \psi= G * \omega\,.
    \label{eq:PoissonFormalSolution}
\end{align}
We approximate the convolution by placing ``point charges'' close to the boundary 
in the region $\mathcal{B}$ where the force (and gorque) acts
\begin{align}
    \psi \approx   \sum_{z_0 \in \mathcal{B}} G(z, z_0)
   \omega(z_0),
    \label{eq:PoissonPointSolution}
\end{align}
where we use Eq.~\eqref{eq:vorticityFinal}  for  computing $\omega(z_0)$. Note that since $\omega(z)$ is only nonzero at the boundary, it is sufficient to evaluate Eq.~\eqref{eq:PoissonPointSolution} at the boundary.

To explicitly compute Eq.~\eqref{eq:PoissonPointSolution}, we define a small boundary region $\mathcal{B}$, which corresponds to the region in which forces act on our dumbbell swimmer (see also Methods section of the main text).
Explicitly, the boundary region $\mathcal{B}$  is defined by an inner ellipse
$\mathcal{E}_{\mathrm{start}}$
and an outer ellipse
$\mathcal{E}_{\mathrm{stop}}$.
Here,
$\mathcal{E}_{\mathrm{start}}$ is characterized by the major half axis $A-a_2$
and minor half axis $B-a_2$, where $a_2$ is the size of the small circle of the dumbbell. 
$\mathcal{E}_{\mathrm{stop}}$ is characterized by the major half axis $A-(2^{1/6} a_1)$
and minor half axis $B-(2^{1/6} a_1)$, where $a_1$ is the size of the large circle of the dumbbell
and the factor $2^{1/6}$ stems from the range of the Weeks--Chandler--Anderson potential used 
to evaluate the forces.

To numerically evaluate Eq.~\eqref{eq:PoissonPointSolution} we further approximate the source term $\omega(z_0)$.
Given a curvature $\kappa$ at the wall we numerically compute the source term $\omega(z_0)$ for a range of distances from the wall 
and average them to obtain $\omega_{\mathrm{avr}}(\kappa)$. 
To find the source term $\omega(z_0)$ at a $z_0 \in \mathcal{B}$ we then compute the local $\kappa$ curvature 
and find a corresponding $\omega_{\mathrm{avr}}(\kappa)$, which is then used to evaluate Eq.~\eqref{eq:PoissonPointSolution}. 
We have to use this procedure since a simple evaluation of $\omega(z_0)$ 
strongly fluctuates and depends on the number of discretization  points.
Thus a simple evaluation of the sum Eq.~\eqref{eq:PoissonPointSolution} does not
give physical results. Using our averaging procedure, however, 
we obtain a smooth approximation of $\omega(z_0)$ 
that does not fluctuate nor depend on the number of discretization points. 

\section{Nonequilibrium flux in circular compartment}\label{sec:circular}
\begin{figure}
    \centering
    \includegraphics[width=0.4\textwidth]{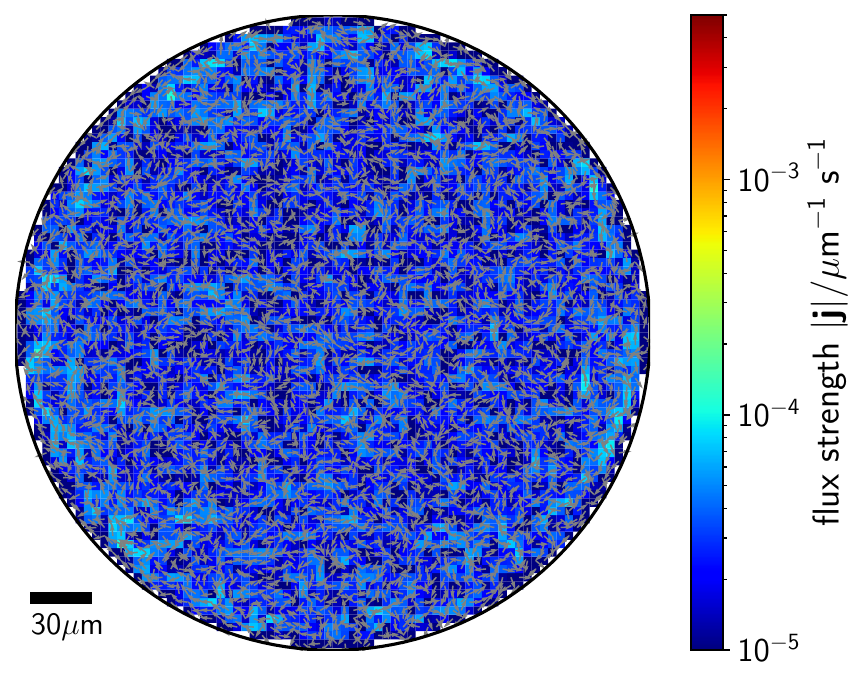}
    \caption{Nonequilibrium fluxes inside a circular chamber, obtained from Brownian 
    dynamics simulations of an asymmetric dumbbell. No directed fluxes are observed.}
    \label{fig:nonequilibrium_flux_circ}
\end{figure}

Figure~\ref{fig:nonequilibrium_flux_circ} shows the nonequilibrium fluxes computed 
from Brownian dynamics simulations of an asymmetric dumbbell (see Eqs.~5-7)
inside a circular compartment.
We do not find any directed fluxes inside the circular chamber. 
This results from the fact that the underlying equations of motion are symmetric in
the polar angle (The effect of activity 
and the corresponding nonequilibrium fluxes in circular chambers can be observed by 
considering the phase space spanned the radial position and the orientation of the 
active particle.).
However, in elliptical chambers the equations of motion are not symmetric in the polar angle.

{
\section{Alternate swimmer geometries}}
\label{sec:other_swimmers}
\begin{figure}
    \centering
    \includegraphics[trim= 5cm 6cm 1.5cm 4cm, clip,width=\textwidth]{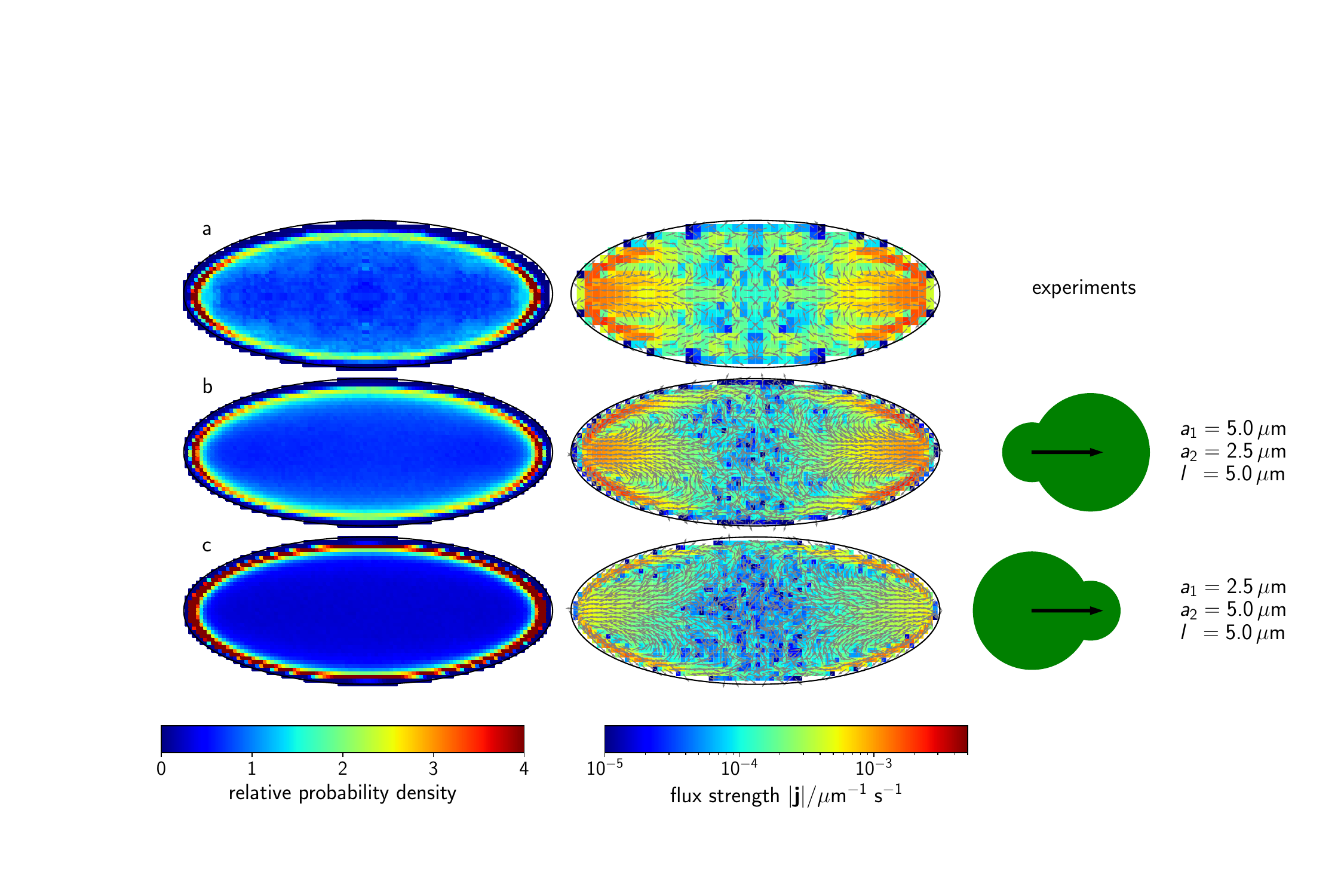}
    \caption{{Comparison of relative probability density (left) and probability fluxes (right) as measured in (a) experiments, (b) from simulations of a swimmer with the fore-aft asymmetry of \emph{Chlamydomonas} (see also Methods), and (c) inverted asymmetry. Results in (b) reproduce the experimental observations most accurately.}}
    \label{fig:other_swimmer_geo}
\end{figure}

{
The emergence of probability fluxes is a direct consequence of active motion and confinement. The choice of geometry of the swimming cell will not qualitatively alter our results, but has quantitative consequences. When direct comparison with experiments is considered, our model (see Methods and Fig.~\ref{fig:dumbbell_illustration}) reproduces the experimental probability fluxes and relative probability density most accurately. As an example, the probability fluxes calculated from experiments and simulations of a swimmer with the fore-aft asymmetry of \emph{Chlamydomonas} and the reversed one are compared in Fig.~\ref{fig:other_swimmer_geo}.\\
}

\section{Other compartment geometries and further comparisons with experiments}\label{sec:shapes}
\begin{figure}
    \centering
    \includegraphics[width=\textwidth]{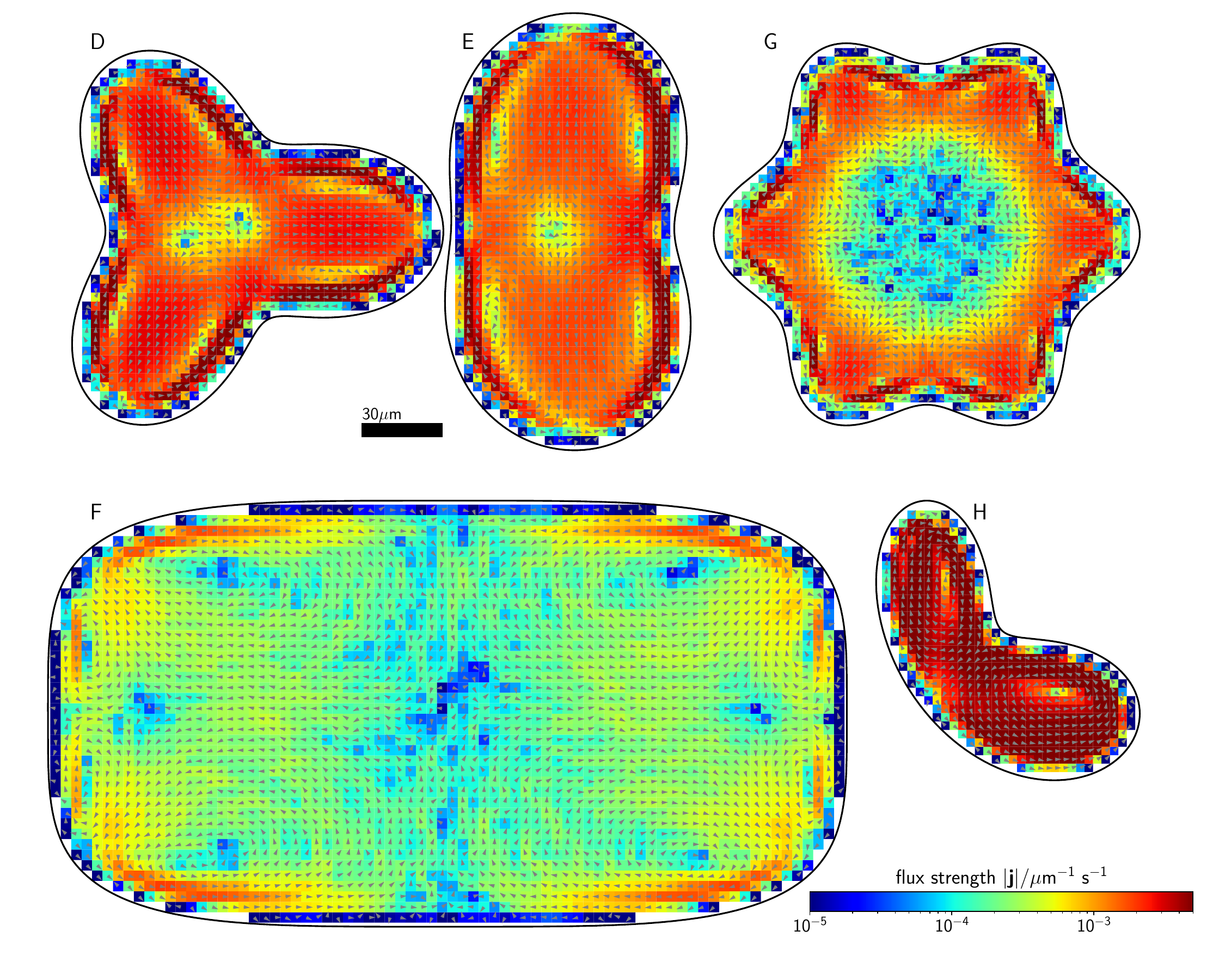}
    \caption{Complexity and topology of flux loops. Complex geometric confinement confirms the topological features of the flux loops. The nonequilibrium fluxes are extracted from our active Brownian dynamics simulations. Fluxes are dominated by gradients of wall curvature. The mathematical formulas of the shown compartments are provided in SI Sec.~\ref{sec:shapes}.}
    \label{fig:compound_si}
\end{figure}

\begin{figure}
    \centering
    \includegraphics[width=\textwidth]{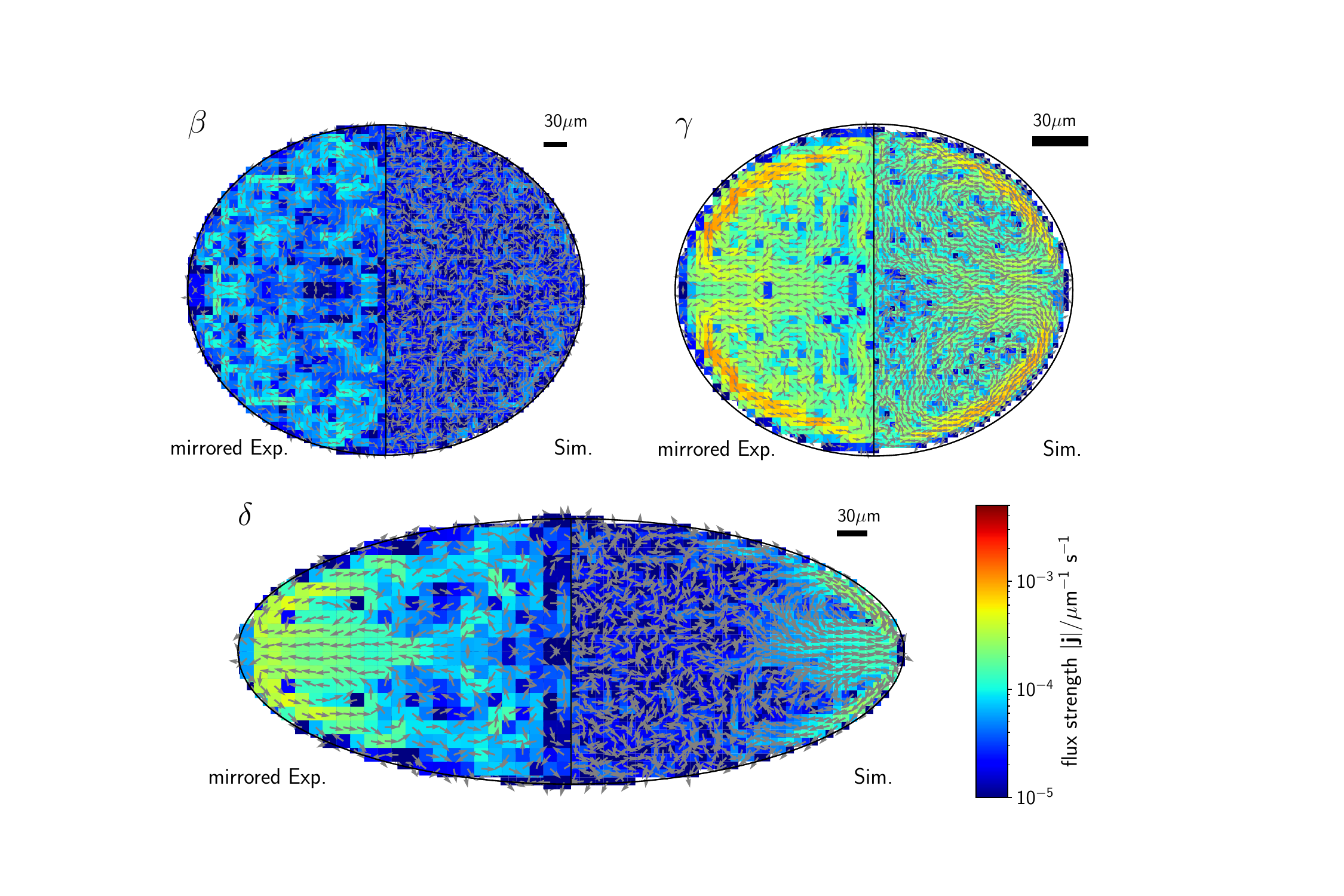}
    \caption{Further comparison of probability fluxes extracted from experiments (left half of each panel) and simulations (right half of each panel). Steady-state nonequilibrium fluxes with arrows indicating their direction and their strength encoded with color. The panel labels $\beta$, $\gamma$, $\delta$ identify the shapes shown in Fig.~4 in the main text.}
    \label{fig:comparison_si}
\end{figure}

The simulation results shown in Fig.~3 and Fig.~\ref{fig:compound_si} where obtained by simulating the dynamics of the introduced active dumbbell model with a confining geometry given by the following polar curves:\\
Shape of compartment a):
\begin{align}\label{eq:potato}
    \frac{r(\theta)}{A}=2+\cos(\theta)+0.1\sin\left(2\theta+\frac{\pi}{3}\right)
\end{align}
Shape of compartment b):
\begin{align}\label{eq:flower}
    \frac{r(\theta)}{A}=1+\sin^2(2\theta)+\frac{1}{2}\sin^2\left(\frac{3}{2}\left[\theta-\frac{\pi}{2}\right]\right)+\frac{1}{2}\sin^2\left(\frac{1}{2}\left[\theta-\frac{\pi}{4}\right]\right)
\end{align}
Shape of compartment d):
\begin{align}\label{eq:lobes3}
    \frac{r(\theta)}{A}=1+\cos^2(1.5\theta)+0.1\sin\left(2\theta+\frac{\pi}{3}\right)
\end{align}
Shape of compartment e):
\begin{align}\label{eq:bean}
    \frac{r(\theta)}{A}=1+\cos^2(\theta)+0.1\sin^2\left(1.5\theta+\frac{\pi}{3}\right)
\end{align}

Shape of compartment g):
\begin{align}\label{eq:six_lobes}
    \frac{r(\theta)}{A}=\frac{9+\cos(6\theta)}{10}
\end{align}

Shape of compartment h):
\begin{align}\label{eq:new_bean}
    \frac{r(\theta)}{A}=\cos^3(\theta)+\sin^5(\theta)
\end{align}
where $A=80\,\upmu$m is the same for all compartments in Eq.~\eqref{eq:potato}-\eqref{eq:new_bean}.\\
Compartments c) and f) are superellipses given by the polar curve:
    \begin{align}
        r(\theta) = \frac{AB}{\left(|B\cos(\theta)|^4+|A\sin(\theta)|^4\right)^{1/4}}
    \end{align}
    with $A=B=80\,\upmu$m for compartment c) and $A=150\,\upmu$m, $B=80\,\upmu$m for compartment f).

We also show additional comparisons between experiments and simulations in Fig.~\ref{fig:comparison_si}.  The panel labels $\beta$, $\gamma$, $\delta$ correspond to the calculations shown in Fig.~4 in the main text. Each panel shows the experimentally obtained probability fluxes (left half of each panel), and the probability fluxes obtained from our simulations of the dumbbell model for \emph{Chlamydomonas} (right half of each panel).

\bibliography{bdb}

\end{document}